\documentclass[12pt,preprint]{aastex}

\shorttitle{ISAAC observation of IRAS 19254-7245}
\shortauthors{G. Risaliti et al.}

\begin{document}

\title{Revealing the AGN in the {\it Superantennae} through L-band spectroscopy\altaffilmark{1}} 

\altaffiltext{1}{Based on observations collected at the European Southern Observatory, Chile (proposal ESO 69.A-0643)}

\author{G. Risaliti\altaffilmark{2,3}, R. Maiolino\altaffilmark{2}, 
A. Marconi\altaffilmark{2}, L. Bassani\altaffilmark{4},
S. Berta\altaffilmark{5}, V. Braito\altaffilmark{5,6}, R. Della Ceca\altaffilmark{6},
A. Franceschini\altaffilmark{5}, M. Salvati\altaffilmark{2}}
%R. Maiolino\altaffilmark{1},
%A. Marconi\altaffilmark{1}    }

\email{risaliti@arcetri.astro.it}

\altaffiltext{2}{INAF - Osservatorio di Arcetri, L.go E. Fermi 5,
Firenze, Italy}
\altaffiltext{3}{Harvard-Smithsonian CfA, 60 Garden st. 
Cambridge, MA}
\altaffiltext{4}{TESRE/CNR, Bologna, Italy}
\altaffiltext{5}{Dipartimento di Astronomia, Universit\`a di Padova, Italy}
%\altaffiltext{6}{INAF - Osservatorio Astrofisico di Padova, Italy}
\altaffiltext{6}{INAF - Osservatorio Astronomico di Brera, Milano, Italy}

\begin{abstract}
We present an L-band spectrum of the Ultraluminous Infrared
Galaxy IRAS 19254-7245 (the {\em Superantennae}), obtained with VLT-ISAAC.
The high signal to noise ratio allows a study of the main spectral features with
unprecedented detail for an extragalactic source. We argue that the main energy source
in the IR is an obscured AGN. This is indicated by the low equivalent width of 
the 3.3~$\mu$m PAH feature, the broad absorption feature at 3.4~$\mu$m, and
the steep continuum at $\lambda>3.7~\mu$m ($f_\lambda\propto\lambda^{2.7}$).
The substructure of the 3.4~$\mu$m absorption feature indicates that the absorption is
due to hydrocarbon chains of 6-7 carbon atoms.  
\end{abstract}


\section{Introduction}

Ultraluminous Infrared galaxies (L$_{IR}>10^{12}~L_\odot$) have been extensively
studied at all wavelengths from the radio to the hard X-rays, in order to 
unveil the energy source - starburst or Active Galactic Nucleus (AGN) - responsible for the huge infrared luminosity.
Determining the AGN role in ULIRGs is fundamental to evaluate the
contribution of accretion to the infrared background, and to have a 
complete view of the population of AGNs in the near Universe.

Several diagnostics have been proposed for this purpose. In the optical, the
presence of high-ionization narrow emission lines is an indication of the presence of 
an AGN. However, it is difficult to estimate the contribution of the AGN
to the total luminosity just from the optical emission lines. 
Hard X-rays are in principle a powerful
tool. The ratio between X-ray and IR emission in starburst-dominated ULIRGs is expected to be
$<10^{-4}$, while in AGN-dominated ULIRGs the ratio is $10^{-3}-10^{-1}$ (Risaliti et al. 
2000). However, 
if the AGN is covered by a column density N$_H$ higher than 10$^{25}$ ~cm$^{-2}$, the 
presence of the AGN and its total luminosity are hard to determine. 
In general, all these methods are useful when they can directly measure the
AGN luminosity, but they fail in determining whether an AGN is present or not in case
of a nondetection.

Mid-infrared spectroscopy provides a powerful way to disentangle the starburst
and AGN contribution. Genzel et al. (1998) used the Polycyclic Aromatic Hydrocarbon (PAH)
molecules emission lines as an indicator of starburst activity in ISO spectra
of the brightest ULIRGs.  Since PAH molecules are 
destroyed by the X-ray radiation emitted by AGNs, a
high equivalent width (EW) of PAH lines is an indicator of the absence of a strong AGN.
Genzel et al. (1998) used the EW of the 
PAH emission feature at $\sim 7.7~\mu$m as main  diagnostics, and concluded
that most ULIRGs are starburst-dominated.
Recently Imanishi \& Dudley (2000) showed that ground-based telescopes can
provide similarly good -or probably better- data for this kind of diagnostics.
In particular, L-band spectroscopy of low redshift ULIRGs allows the direct 
measurement of the $3.3~\mu$m PAH feature (indicator of starburst activity, similarly
to the 7.7~$\mu$m feature) and of the
carbonaceous dust absorption dip at $\sim 3.4~\mu$m (indication of an absorbed point
source like an AGN). Most importantly, the L-band is wide enough for a correct
estimate of the continuum level, needed to measure both the
equivalent width of the PAH feature and the detection of the carbon dip. 
In this way Imanishi et al. (2001) and Imanishi \& Maloney (2003)
clearly discovered signatures of an AGN dominating
the energy output in UGC 5101, which was previously classified as
starburst-dominated using ISO spectra only. 

The ULIRG IRAS 19254-7245 (the {\it Superantennae}, Mirabel, Lutz \& Maza 1991) 
is optically classified 
as a Seyfert 2. ISO mid-infrared spectroscopy indicates the presence of an AGN in IRAS 19254-7245,
but it does not provide a clear indication
on which is the dominant energy source, the EW of the 7.7~$\mu$m PAH feature being
intermediate between typical AGN and typical starburst values (Genzel et al. 1998).
Recently Charmandaris et al. (2002) classified the source as AGN-dominated, based
on the same ISO data.

XMM-Newton data in the 2-10 keV band suggests that IRAS 19254-7245
is AGN-dominated (Braito et al. 2002). The X-ray spectrum suggests that the source
is Compton-thick, i.e. the absorbing column density is higher than 10$^{24}$~cm$^{-2}$.
The total luminosity of the AGN is estimated to be of the order or higher than
10$^{44}$~ergs~s$^{-1}$.
%, but is starburst-dominated according to ISO mid-infrared
%spectroscopy (Genzel et al. 1998). However, an XMM-Newton observation in the
%2-10 keV band suggests that it is AGN-dominated (Braito et al. 2003). The same
%conclusion has been reached by Vanzi et al. (2001) based on optical and near-IR
%spectroscopy. 
Berta et al. (2003) fitted the SED of IRAS 19254-7245 from the U to the mm-band,
with a starburst+AGN model, estimating a contribution of the AGN to the
bolometric luminosity of 40-50\%.

Here we present an L-band spectrum of IRAS 19254-7245, obtained with the
instrument ISAAC on the VLT as part of a mini-survey of
bright ULIRGs. Thanks to the superb quality of both the
telescope and the instrument, this is probably the best L-band spectrum of
a ULIRG ever published, and shows the potentiality of ground-based L-band
spectroscopy in the study of active and star forming galaxies.

\section{Data reduction and analysis}

IRAS 19254-7245 is a system consisting of two colliding spiral 
galaxies, at z=0.062. The separation of 
the two nuclei is 8.5'' ($\sim10$~kpc\footnote{We assume H$_0$=75 km s$^{-1}$
Mpc$^{-1}$.}). The southern galaxy is classified as a Seyfert 2 (Mirabel et al.
1991), while the northern galaxy shows an optical spectrum typical of
starbursts (Colina, Lipari \& Macchetto 1991). At 2~$\mu$m, the southern galaxy is 
brighter by $\sim$1 magnitude than the northern galaxy (Duc, Mirabel \& Maza 1997).
Images at  10~$\mu$m show that the mid-IR emission is concentrated in the
nuclei, and the southern one is more than $\sim$5 times brighter than the northern one.
Finally, ISOCAM observations (Charmandaris et al. 2002) 
show that more than 90\% of the 5-20~$\mu$m
emission is due to the southern source.

We performed 2 observations of IRAS 19254-7245 with ISAAC-VLT, on June 02, 2002, and on June
3, 2002. Each observation was 1 hour long. The two nights were photometric, and
the seeing was around 1''.
In the image acquired before the spectroscopic observation, we detected the
two nuclei and we oriented the 1'' slit in order to obtain spectra of both.
However, the northern nucleus turned out to be too faint to obtain a useful
spectrum. Therefore, we will not discuss it further.

The spectroscopic observations have been performed in ``chopping'' mode, 
with
single exposures of 0.56 sec. The data were merged,
flat-fielded and sky-subtracted using standard procedures in the IRAF
package. A spectrum of the spectrophotometric standard star HIP 183 (B4III, L=5.5,
T$_{eff}$=15,800 k) was acquired in the
same way for both observations, and used to obtain the instrumental
response, in order to correct the source spectrum. 

Since the calibration lamp lines were too faint to be useful,
the wavelength calibration was performed using the nominal instrumental range
and the wavelength of the carbon absorption features (as described in the next
Section). 
Throughout this paper we always refer to rest frame wavelengths, unless
otherwise stated.

The absolute flux
calibration has been obtained analyzing the profile of the star along the
slit, and estimating the fraction of flux inside the slit, assuming a perfect 
centering. We estimate that this procedure  has an error of the order of
10\%.
We checked that the two final spectra were consistent, and we finally merged
the results of the two observations. 
The final spectrum, rebinned by a factor of 5, is shown in Fig. 1.  
The error bars are estimated from the Poissonian noise in the sky counts (which are 
by far the dominant source of noise).

%\clearpage

\begin{figure}
\plotone{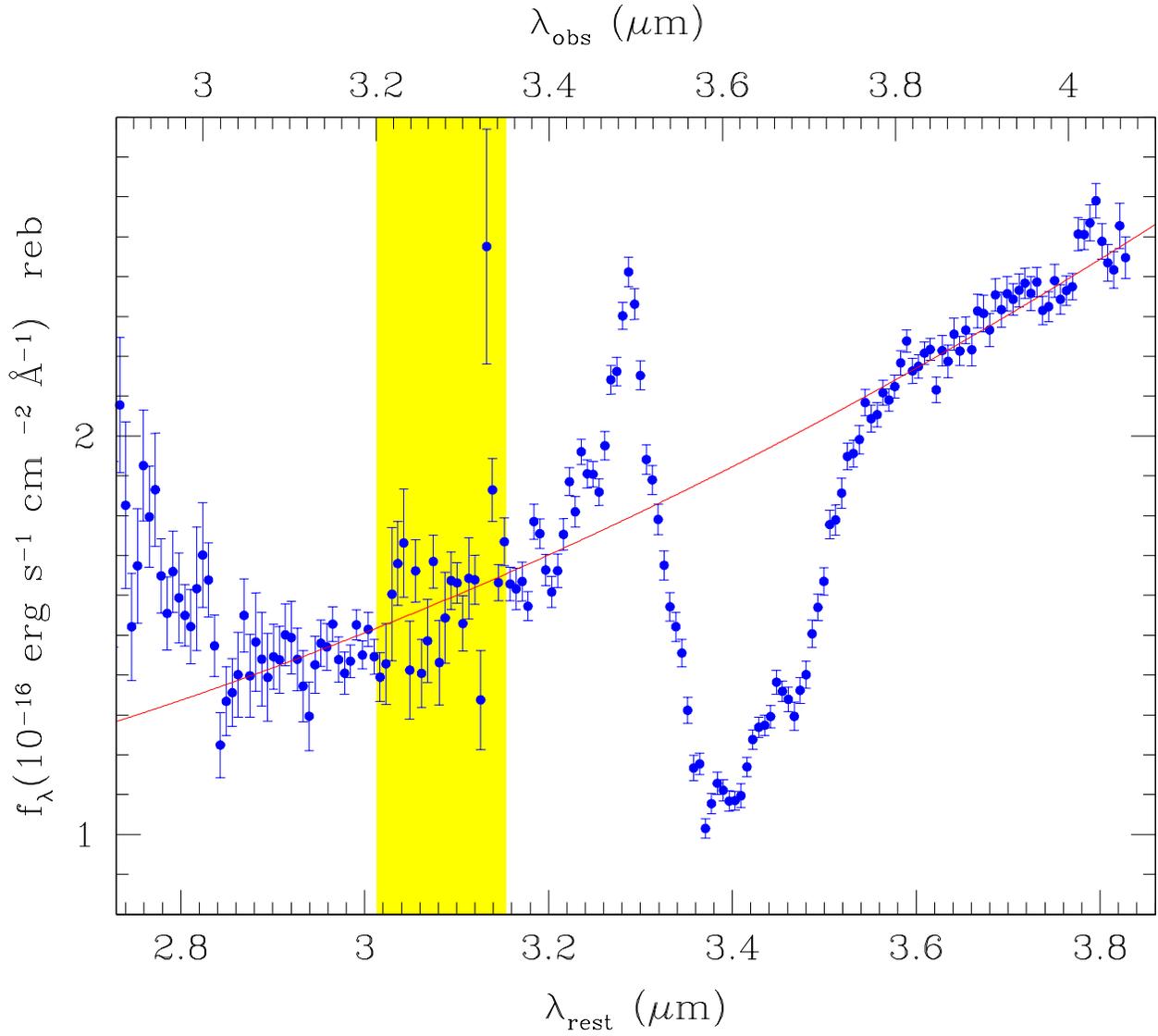}
\figcaption{\footnotesize{
Flux-calibrated spectrum of IRAS 19254-7245. In the observed wavelength
range 3.2-3.35~$\mu$m (shaded region) the atmospheric transmission is poor.
The solid line is an adopted continuum.}}
\end{figure}
%\clearpage

\section{Discussion}

Previous optical and infrared observations  indicated the
presence of an AGN in the nucleus of IRAS 19254-7245, but the 
relative importance of the AGN and starburst components was not
clear: the 7.7~$\mu$m PAH feature detected by ISO (Genzel et al. 1998)
has an intermediate strength (S=0.8, where S is the ratio between 
the peak 7.7~$\mu$m flux and the continuum at the same wavelength) with
respect to pure starbursts and pure AGN (S$\sim$3.6 and S$\sim$0.04
respectively, Genzel et al. 1998).
However, the determination of the continuum level is highly uncertain in
the ISOPHOT spectra, due to the presence of absorption features and
the poor signal-to-noise ratio.
%Based on the same data, Charmandaris et al. (2001) concludes that the
%source is probably dominated by the AGN emission. 
In the optical and near-IR, the presence of an AGN is clearly revealed from the
high [OIII]/H$\beta$ ratio and from the presence of strong coronal lines
(Vanzi et al. 2002).

The ISAAC spectrum presented here is probably the highest
S/N L-band spectrum of a ULIRG so far. Thanks to the high signal
and to the moderately high spectral resolution, the starburst and
AGN indicators can be studied with unprecedented detail.

The signatures of the AGN activity in the L-band spectrum of IRAS 19254-7245 are
quite clear: 1) the deep absorption at $\sim3.4~\mu$m (rest frame) strongly
suggesting the presence of a point source behind a screen of dusty gas;
2) the PAH feature at 3.3~$\mu$m with a much lower equivalent width 
than typical starburst-dominated sources; 3) the rather steep continuum slope above
$\sim3~\mu$m, suggesting the presence of warm, AGN-heated
dust. We discuss each of these points in the following.

\subsection{The absorption feature at 3.4~$\mu$m}

The broad absorption feature at $\sim3.4~\mu$m (rest frame) is present
in the spectra of many AGNs and ULIRGs (Imanishi \& Dudley 2000, Imanishi
2000). We clearly detected such feature in the spectrum of
IRAS~19254-7245, with an optical depth $\tau\sim0.8$.
The absorption is believed to be due to C-H stretching vibration
in hydrocarbon dust grains (Sandford et al. 1991 and references therein).
%A strong such feature is considered a reliable indicator of the presence of an active
%nucleus.
Imanishi \& Maloney (2003) showed that an optical depth of this feature
$\tau_{3.4}$ higher than $\sim0.2$ requires a centrally-concentrated source,
i.e. an AGN, unless dust absorption in the host galaxy is significant. 
%, both for theoretical and observational reasons:\\
%- A deep continuum absorption requires a dense ({\rm quanto denso? possiamo
%mettere qualche numero?}) dusty medium fully obscuring the 
%energy source. This is what is expected in obscured AGN. On the other side, a
%starburst emitting $\sim10^{45}$~erg~s$^{-1}$ or more would have a linear
%size of at least 1 kpc ({\rm ???? ragionevole??? ????} ) and it is unlikely
%that such a large region is completely covered by this dusty absorber.\\
Indeed, absorption features at 3.4~$\mu$m are commonly observed in AGNs
(Imanishi 2000) while they are never found in galaxies known to be dominated by
starburst. 

Since the same dust grains are present in the Galactic interstellar medium,
the most detailed studies of the 3.4~$\mu$m absorption have been done on Galactic center
sources (Sandford et al. 1991, Pendelton et al. 1994). 
In these cases the high statistics available allows to resolve the
substructure of the absorption feature in at least two major components,
due to -CH$_2$ and -CH$_3$ groups, respectively. 
%The ISAAC spectrum of IRAS 19254-7245 is the first of an extragalactic source
%where this sub-structure is clearly identified.
In Fig. 2b we plot the ratio between the observed spectrum and the continuum
shown in Fig. 1, obtained with a polynomial fit of the regions with
observed wavelengths in the ranges 2.9-3.1~$\mu$m, 3.3-3.5~$\mu$m, 3.8-4.1~$\mu$m.
A similar spectrum of the Galactic center source IRS7 from Pendelton et al.
1994 is plotted in Fig. 2a. The similarity between the two profiles is
remarkable, and suggests a similar composition of the carbonaceous dust
grains. A deblending of the single absorption components is quite uncertain:
at least three components are present at rest frame wavelengths of 3.38,
3.42 and 3.485~$\mu$m (Fig. 2). Moreover, the absorbed continuum 
is blended with the long wavelength tail of the PAH emission feature at 3.3~$\mu$m.

A qualitative analysis is possible comparing our data with those
of the source IRS 7. Following Sandford et al. (1991) the 
absorption features at 3.38~$\mu$m and at 3.42~$\mu$m are due to
-CH$_3$ and -CH$_2$ groups, respectively, in saturated
aliphatic hydrocarbons. The ratio between the two absorption peaks is
an indication of the average length of hydrocarbon chains\footnote{The
structure of a saturated hydrocarbon chain is CH$_3$-(CH$_2)_n$-CH$_3$,
therefore the ratio between the strength of -CH$_3$ and -CH$_2$ absorption features 
is a direct measure of $n$.}.
Since the relative strength of the two features
in IRAS 19254-7245 is compatible within the errors with
those in IRS~7, the average ratio between -CH$_2$ and -CH$_3$ groups 
is the same, i.e. $\sim 2-2.5$. This
corresponds to relatively complex molecules made of 6-7 carbon atoms.
%like hexane.
The third absorption feature is due to the same C-H stretching vibrations,
when the -CH$_2$ and CH$_3$ groups are perturbed by an electronegative group,
such as -C$\equiv$N, -OH, or aromatic chains. The higher depth of this
feature in  IRAS 19254-7245 with respect to the Galactic source IRS~7
indicates that the absorbing medium in IRAS 19254-7245 is rich of 
organic molecules with such electronegative groups (see Fig. 7 in Sandford et
al. 1991).

%\clearpage

\begin{figure}
\epsscale{0.8}
\plotone{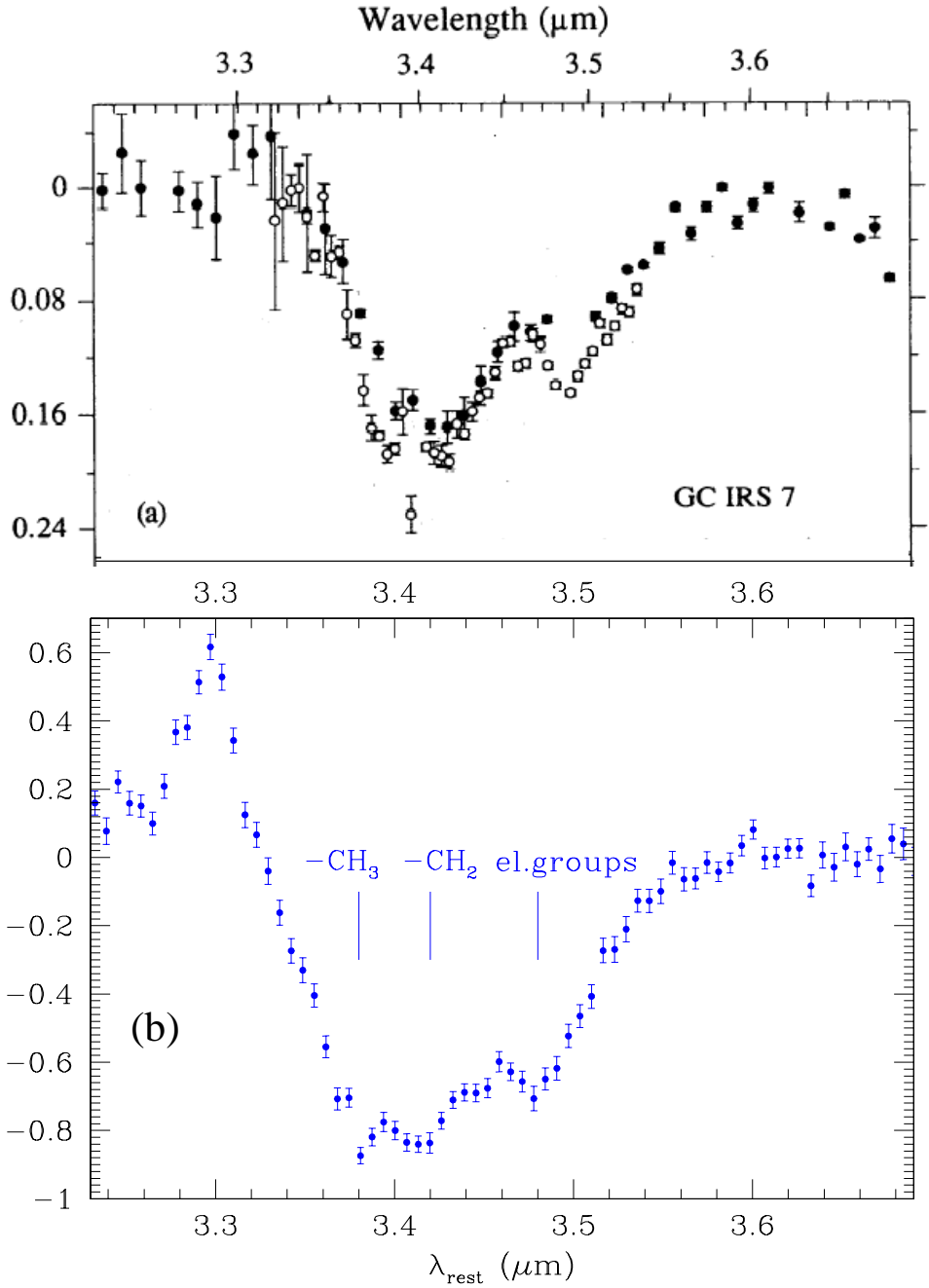}
\figcaption{\footnotesize{(a): Continuum-subtracted L-band spectrum of the Galactic
center source IRS~7, from Pendelton et al. 1994. (b): The same for IRAS
19254-7245. The similarity in the structure of the absorption feature
indicates a common origin, i.e.vibrational transitions of -CH$_3$ and -CH$_2$
groups in hydrocarbon grains.
}}
\end{figure}

%\clearpage

\subsection{The emission feature at 3.3~$\mu$m}
The broad emission feature at 3.3~$\mu$m is due to Polycyclic Aromathic Hydrocarbon
molecules. 
It has been empirically shown that this emission feature is weak or
absent in AGNs, while is strong in starbursts (Roche et al.
1991).
This is probably due to the destruction of PAH molecules by the strong X-ray 
and EUV emission
of AGNs, and to the dilution of the continuum stellar component . As a consequence,  a low
equivalent width (EW) of the 3.3~$\mu$m emission feature is an indicator of 
AGN emission. Typical EW for starburst galaxies are in the range 
1000-1500~\AA~(Moorwood 1986, Imanishi \& Dudley 2000), while in AGN-dominated sources
EW $\leq$100-300~\AA~(Imanishi 2002). 
The estimated value for IRAS 19254-7245 is EW=120$\pm10$~\AA, suggesting that
the emission is dominated by the active nucleus. 
%The error on this estimate due to aperture effects
%is estimated to be smaller than 10\%.

We emphasize that the broad spectral range is crucial for a correct estimate of
the EW. In particular, if the continuum at rest-frame wavelengths $\lambda>3.6~\mu$m is
not available, the line flux is easily over-estimated,
while the continuum  at 3.4~$\mu$m is under-estimated because of the 
broad 3.4~$\mu$m absorption feature. To illustrate this fact, 
we tried to fit the rest-frame 3.0-3.7~$\mu$m
spectrum with a simple power law and an emission line, and we obtained an
equivalent width of 700~\AA, in the range typical of starburst-dominated
sources. 
Another possible source of uncertainty in the measurement of the 3.3~$\mu$m feature
EW is the overlap with the 3.4~$\mu$m absorption feature.
However, the 3.3~$\mu$m emission line appears to be symmetric with respect its
peak wavelength (which is exactly at 3.3~$\mu$m), and the absorption feature in the 
comparison source in Fig. 2a does not extend below $\sim3.34~\mu$m.
We conclude that the 3.3~$\mu$m PAH feature is not significantly affected
by the absorption at 3.4~$\mu$m.
%As a result, the EW of the
%PAH emission can be over-estimated by a factor 7-10, up to values as high as
%those typical of starburst-dominated sources.

It is possible to use the measurement of the 3.3 PAH line to estimate
the AGN/starburst contribution to the infrared emission.
An interesting indicator 
%of the starburst/AGN contribution to the
%infrared emission 
is the ratio between the flux of the 3.3~$\mu$m feature, 
$F_{3.3}$, and the total far infrared flux,F$_{IR}$, 
as estimated from the four IRAS filters
(Sanders \& Mirabel 1996). The typical value for starburst galaxies is 
$R\sim10^{-3}$, while in AGN-dominated sources $R$ is more than 1-2 orders of magnitude
lower (Imanishi 2002). For IRAS 19254-7245 we have F$_{IR}=5.3\times10^{-10}$
erg~cm$^{-2}$~s$^{-1}$ and F$_{3.3}=2.1\times10^{-14}$ erg~cm$^{-2}$~s$^{-1}$.
As a consequence, $R=3.9\times10^{-5}$. 
This indicator suggests that the detected starburst can account for only a small 
fraction of the infrared luminosity of IRAS~19254-7245.
%Again, this indicator confirms that
%the dominant energy source in IRAS 19254-7245 is an active nucleus.

A direct way to obtain a rough quantitative estimate of the AGN contribution to the
infrared luminosity is to subtract a starburst template from the L-band spectrum,
and then compare the L-band flux $(\lambda f_\lambda)_{3\mu m}$
with the total Far Infrared flux $F_{FIR}$ as measured
by IRAS. We expect that in case of AGN dominance these two fluxes are $\sim$ equal.
We assumed that the PAH emission is entirely due to the starburst, and that the starburst 
continuum is reproduced by a flat ($f_\lambda=$const) spectrum normalized in order
to have EW$_{3.3\mu m}$=1000~\AA (in agreement with starburst spectra in Imanishi \& Dudley 2000).
The resulting ``pure AGN'' spectrum is steeper ($\Gamma\sim3)$, with 
%a 3.4~$\mu$m optical depth
$\tau_{3.4 \mu m}=0.9$ and 
%a 3~$\mu$m flux 
$(\lambda f_\lambda)_{3\mu m}=6\times10^{-12}$~erg~s$^{-1}$~cm$^{-2}$.
Assuming the Galactic extinction curve of Pendelton et al. 1994 ($A_V\sim140-250\tau_{3.4 \mu m}$, 
$A_L\sim0.06~A_V$, where $A_L$ is the extinction in the L-band), we obtain $A_L\sim8-14$.
Correcting the observed 3~$\mu$m flux for this extinction we end up with 
$(\lambda f_\lambda)_{3\mu m}=2-4500\times F_{FIR}$. This result is an indication that the 
AGN is dominant in the IR, and that the extinction curve in the {\em Superantennae} 
is probably quite different from the Galactic
one.
%: in order to have $(\lambda f_\lambda)_{3\mu m} = F_{FIR}$ the L-band extinction should be 
%$A_V=6$.
%Further details on this issue will be discussed in a forthcoming paper (Risaliti
%et al. 2003, in preparation).
%\clearpage

\begin{figure}
\plotone{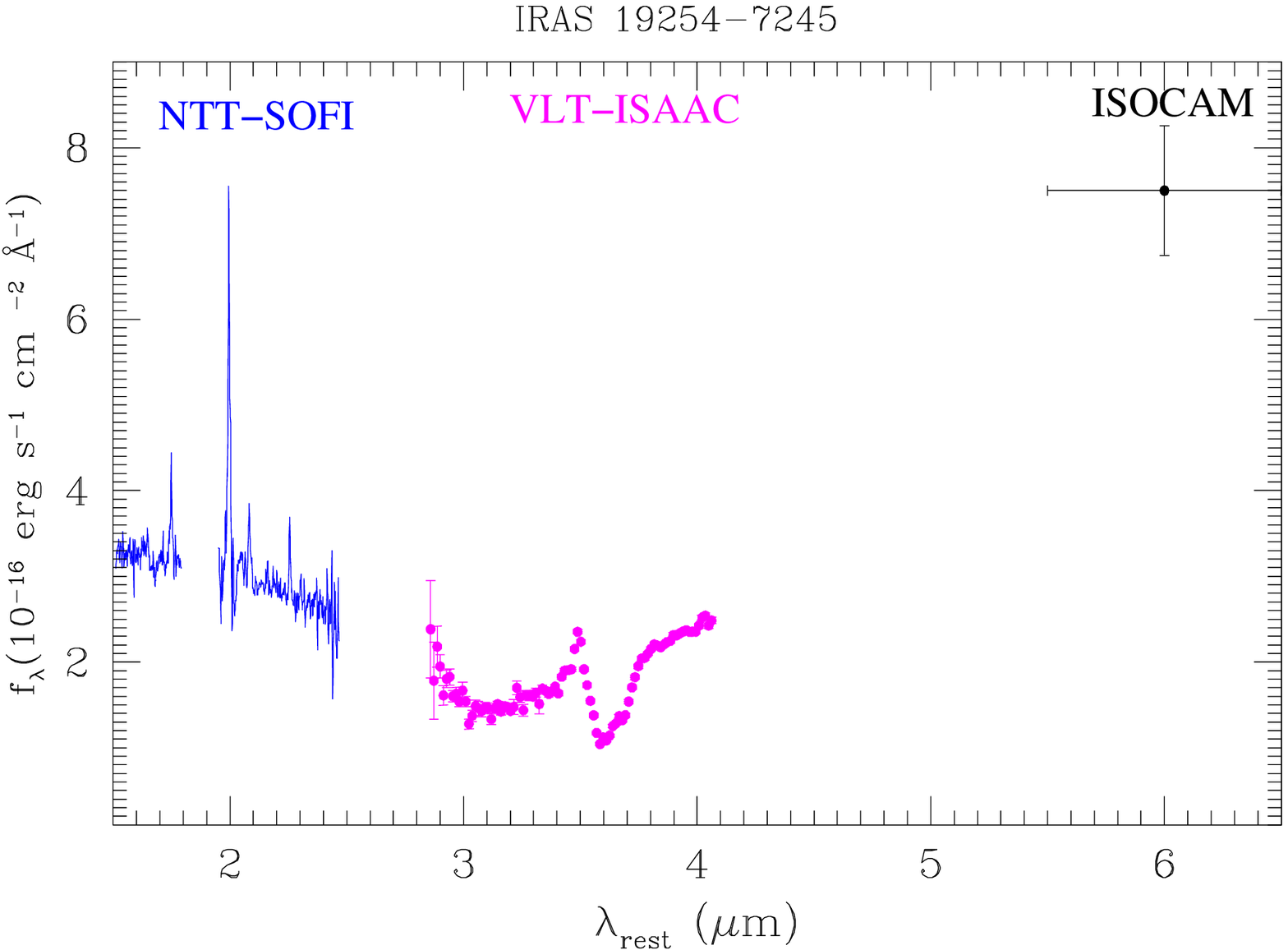}
\figcaption{\footnotesize{1-6~$\mu$m spectrum of IRAS 19254-7245.
}}
\end{figure}

%\clearpage

\subsection{Continuum slope}
The slope of the continuum at wavelengths $\lambda>3.7~\mu$m is rather steep
($f_\lambda\propto\lambda^{2.7}$). This suggests a strong emission by 
warm dust, typical of AGN spectra (Granato, Danese \& Francecshini 1997). 
Interestingly, analyzing the 3-4~$\mu$m spectra of ULIRGs available in the
literature 
%(in particular in Imanishi \& Dudley 2000) 
we note that a positive
slope in the 3.5-4.0~$\mu$m continuum is only found in AGN-dominated sources.
The opposite is not true: a few sources exist (MKN 231, NGC 6240) with
negative 3.5-4.0~$\mu$m slope, which are known to host a powerful AGN from
observations at other wavelengths. This issue will be discussed in further
detail in a forthcoming paper 
%discussing ISAAC observations of 6 more ULIRGs
(Risaliti et al. 2003, in preparation).
At $\lambda<3.1~\mu$m an inversion 
of the slope is clearly seen. In Fig. 3 we plot the L-band spectrum 
with the near-IR spectrum obtained at the NTT (Vanzi et al. 2002)
and the ISOCAM point at 6~$\mu$m (Laurent et al. 2000). 
The extrapolation of the
continuum matches well both the K spectrum and the 6~$\mu$m point. This shows
that the three observations are well cross-calibrated, and our flux
calibration error is lower than $\sim$5\%.

The minimum in the L-band spectrum, at $\sim$3.1~$\mu$m is the absolute
minimum in the optical-infrared spectrum of IRAS 19254-7245. At shorter 
wavelengths the emission is dominated by radiation from 
the host galaxy, since the AGN is highly obscured-, while at longer
wavelengths the radiation is mainly due to the reprocessing of the AGN
direct emission. Imanishi \& Maloney (2003) detected a broad absorption feature 
at 3.1~$\mu$m in several ULIRGs, due to ice-covered dust grains. In principle we
cannot exclude that the minimum observed at 3.1~$\mu$m is partly due to this effect.
However, we note that if a significant fraction of carbonaceous dust grains are covered
by ice, they do not contribute to the 3.4~$\mu$m absorption. This would imply that
the optical extinction $A_V$ for a given $\tau_{3.4 \mu m}$ is even higher than
that assumed in Sect. 3.2. If this is the case, the correction to the
observed 3~$\mu$m continuum would also be higher, worsening the problem
of the too high 3~$\mu$m flux compared with the FIR emission.

\section{Conclusions}

We presented a high signal-to-noise L-band spectrum of
the ultra-luminous infrared galaxy IRAS 19254-7245.
The signatures of a powerful AGN dominating the energy output in the infrared
are clear:\\
$\bullet$ The broad absorption feature at $\sim3.4~\mu$m, suggests
the presence of a powerful point source. The substructure of this feature shows
that the absorption is due to hydrocarbon molecules consisting of chains of 6-7 carbon atoms,
rich in electronegative groups, as -CN, -OH.\\
$\bullet$ The EW of the PAH emission feature at 3.3~$\mu$m, when the continuum is
correctly estimated taking into account the broad absorption at 3.4~$\mu$m, is
typical of AGN-dominated sources.\\
$\bullet$ The continuum at $\lambda>3.7~\mu$m is rather steep, indicating the
presence of hot dust.\\
The above conclusions show the potential of high signal-to-noise L-band
spectroscopic observations with VLT-ISAAC. We are now completing
a work in which we extend the present study to a sample of 6 more bright
ULIRGs of the Genzel et al. (1998) sample (Risaliti et al. 2003, in
preparation).

\acknowledgements
We gratefully acknowledge Dr. D. Lutz for fruitful discussions, and 
the anonymous referee for suggestions which significantly
improved this Letter.
This work has been partially supported by the Italian Ministry of Research (MIUR)
and the Italian Space Agency (ASI).


\begin{thebibliography}{}
\bibitem[]{384} Berta, S., et al. 2003, A\&A, submitted
\bibitem[]{385} Braito, V., et al. 2003, A\&A, 398, 107
%\bibitem[]{385} Bridger, 
%A., Wright, G.~S., \& Geballe, T.~R.\ 1994, ASSL Vol.~190: Astronomy with 
%Arrays, The Next Generation, 537 
\bibitem[Colina, Lipari, \& Macchetto(1991)]{1991ApJ...379..113C} Colina, 
L., Lipari, S., \& Macchetto, F.\ 1991, \apj, 379, 113 
\bibitem[]{386} Charmandaris, V., et al. 2002, A\&A, 391, 429 
\bibitem[Duc, Mirabel, \& Maza(1997)]{1997A&AS..124..533D} Duc, P.-A., 
Mirabel, I.~F., \& Maza, J.\ 1997, \aaps, 124, 533 
\bibitem[]{387} Genzel, R., et al. 1998, ApJ, 498, 579
\bibitem[]{388} Granato, G.L., Danese, L., \& Franceschini, A.
1997, ApJ, 486, 147
%\bibitem[]{388} Imanishi, M., Terada, 
%H., Sugiyama, K., Motohara, K., Goto, M., \& Maihara, T.\ 1997, \pasj, 49, 69 
\bibitem[]{389} Imanishi, M. 2000, MNRAS, 319, 331
\bibitem[]{390} Imanishi, M., \& Dudley, C.C. 2000, ApJ, 545, 701
\bibitem[]{391} Imanishi, 
M., Dudley, C.~C., \& Maloney, P.~R.\ 2001, \apjl, 558, L93 
\bibitem[]{392} Imanishi, M. 2002, ApJ, 569, 44
\bibitem[]{393} Imanishi, M.~\& 
Maloney, P.~R.\ 2003, \apj, 588, 165 
\bibitem[]{394} Laurent, O., et al. 2000, A\&A, 359, 887
\bibitem[]{395} Mirabel, I.F., Lutz, D., Maza, J., 1991, A\&A, 243, 367
\bibitem[Moorwood(1986)]{1986A&A...166....4M} Moorwood, A.~F.~M.\ 1986, 
\aap, 166, 4 
\bibitem[]{396} Pendleton, Y.J., Sandford, S.A., Allamandola, L.J.,
Tielens, A.G.G.M., \& Sellgren, K. 1994, ApJ, 437, 683
\bibitem[]{397} Risaliti, G., Gilli, R., Maiolino, R., \& Salvati, M.
2000, A\&A, 357, 13
\bibitem[]{398} Roche, P. F., Aitken, D. K., Smith, C. H., \& Ward, M.
1991, MNRAS, 248, 606
\bibitem[Sanders \& Mirabel(1996)]{1996ARA&A..34..749S} Sanders, D.~B.~\& 
Mirabel, I.~F.\ 1996, \araa, 34, 749 
\bibitem[]{399} Sandford, S.A., Allamandola, L.J., Tielens, A.G.G.M., 
Sellgren, K., Tapia, M., \& Pendleton, Y.J. 1991, ApJ, 371, 607
\bibitem[]{401} Vanzi, L., Bagnulo, S., Le Floc'h, E., Maiolino, R.,
Pompei, E., \& Walsh, W. 2002, A\&A, 386, 464 
\end{thebibliography}
\end{document}